\documentclass[reprint,amsmath,amssymb,aps,prl,superscriptaddress]{revtex4-1}

\usepackage{graphicx}% Include figure files
\usepackage{dcolumn}% Align table columns on decimal point
\usepackage{bm}% bold math
\usepackage{color}
\usepackage{ulem}

\begin{document}

\title{Cavitation nuclei regeneration in water-particle suspension}

\author{Adrien Bussonni\`ere$^{1,2}$, Qingxia Liu$^2$, Peichun Amy Tsai}

\affiliation{Department of Mechanical Engineering;~$^2$Department of Chemical and Materials Engineering, University of Alberta, Edmonton, AB, T6G 1H9, Canada.}

\date{\today}

\begin{abstract}
Bubble nucleation in water induced by boiling, gas supersaturation or cavitation usually originates from pre-existing gas cavities trapped into solid defects. Even though the destabilization of such gas pockets, called nuclei, has been extensively studied, little is known on the nuclei dynamic. Here, nuclei of water-particle suspensions are excited by acoustic cavitation, and their dynamic is investigated by monitoring the cavitation probability over several thousand pulses. A stable and reproducible cavitation probability emerges after a few thousand pulses and depends on particle concentration, hydrophobicity, and dissolved gas content. Our observations indicate that a stable nuclei distribution is reached at a later-time, different from previously reported nuclei depletion in early-time. This apparent paradox is elucidated by varying the excitation rate, where the cavitation activity increases with the repetition period, indicating that the nuclei depletion is balanced by spontaneous nucleation or growth of nuclei. A model of this self-supporting generation of nuclei suggests an origin from dissolved gas adsorption on surfaces. The method developed can be utilized to further understand the spontaneous formation and distribution of nano-sized bubbles on heterogeneous surfaces. 
%\begin{description}

%\item[PACS numbers]

%\end{description}
\end{abstract}

\maketitle
The formation of bubbles in water by either cavitation (through pressure reduction), boiling (via temperature increase) or gas supersaturation experimentally happens at much smaller values than those predicted by the classical nucleation theory \cite{Caupin2006,Dhir1998,Jones1999}. This discrepancy has been rationalized by the presence of small gas pockets, named Harvey nuclei \cite{Harvey1944}, trapped into the surface defects of a wall or floating particles, and bubble formation originates from the loss of stability of such nuclei. Prediction of the critical value for nuclei destabilization has been extensively studied over several years \cite{Jones1999,Kocamustafaogullari1983,Wang1993,Basu2002} and led to the so-called crevice model in cavitation \cite{Apfel1970,Crum1979,Atchley1989}. This theory was found in excellent agreement with experimental data using geometrically-controlled cavitation nuclei \cite{Borkent2009}. However, fewer studies have been dedicated to the nuclei dynamic and origin \cite{Wang2017}, despite playing a pivotal role in bubble formation. Nuclei are generally assumed to form during solid immersion \cite{Bankoff1958,Atchley1989} and are known to evolve with liquid properties such as the dissolved gas content, temperature or static pressure \cite{Strasberg1959,Apfel1970,Crum1979}. Recently, Borkent et al. \cite{Borkent2007} showed a decrease of the cavitation activity of a water-particle suspension over consecutive acoustic pulses, implying a depletion of the nuclei population, i.e., a nucleus acting as a cavitation site only once. Nuclei deactivation by acoustic cavitation was further studied and confirmed using controlled pits \cite{Borkent2009}. This decrease suggests that the nuclei population is finite, only determined by an initial state (dependent on the mixture properties and history) and can be controlled by pre-cavitating a solution. 

To further verify these suggested behaviors, which could lead to a control of bubble formation, we investigate the cavitation of water-particle mixture excited by acoustic pulses over several hours, i.e., several tens of thousand pulses. We discover that cavitation after such long time does not vanish but instead reaches a stable regime. This stable long-time cavitation is found to originate from a balance between the acoustic nuclei depletion and a spontaneous nuclei regeneration.

\begin{figure*}
\begin{center}
    \includegraphics[width = 17 cm]{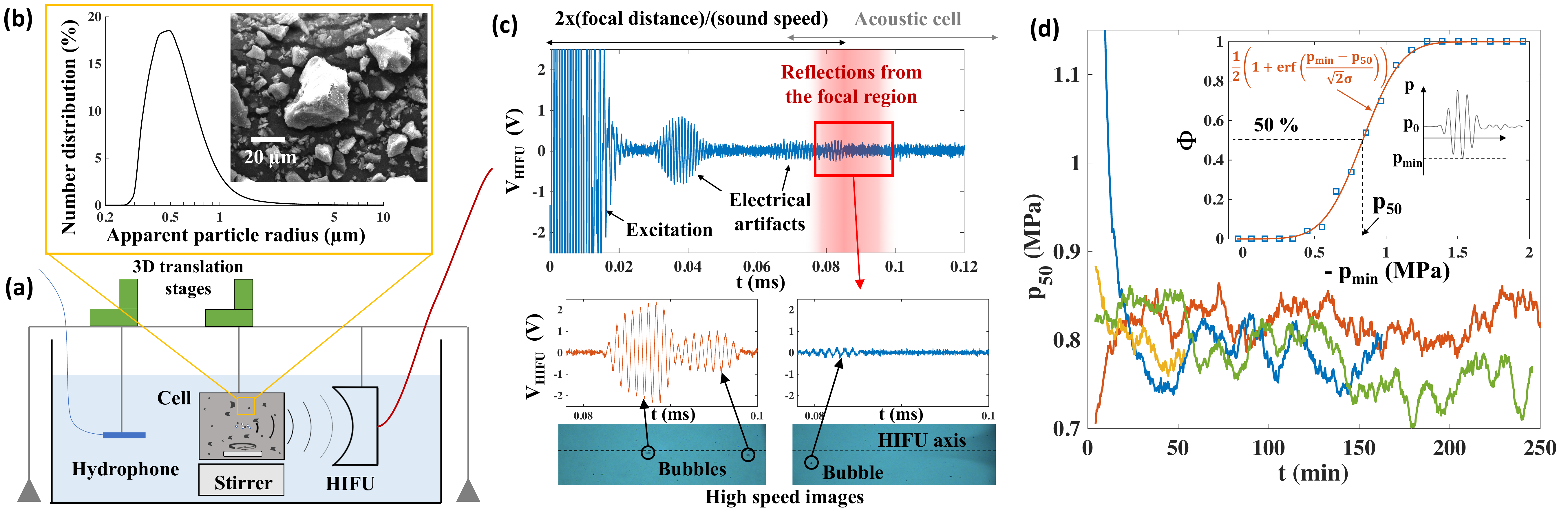}
\caption{\label{FIG1}{Sketch of the cavitation experiment. a) Pressure pulses are focused by a HIFU inside a cell containing a ground silica suspension. The pressure field is calibrated with a hydrophone. b) Size distribution and SEM picture of the particles. c) Voltage of the HIFU showing the echo signals induced by the nucleated bubbles. Two examples of the echo signals are depicted with synchronized images showing the nucleated bubbles. d) Time evolution of the critical pressure $p_{50}$ for four different suspensions of 1 g/l. The inset depicts the method of extracting $p_{50}$ from the cavitation probability curve $\Phi$ in term of the minimum pressure at the focal point, $p_\text{min}$.}}
\end{center}
\end{figure*}

In this Letter, cavitation of a particle-water mixture is experimentally studied for the first time using a probabilistic approach, better suited to probe systems involving unknown parameters, here the nuclei distribution. Since impurities are extremely difficult to control, even with well-calibrated particles \cite{Marshall2003}, we choose ground silica particles with random shapes (see Fig.~\ref{FIG1}a-b) in a large quantity (typically $10^7$ particle/ml) to promote the emergence of a reproducible nuclei distribution. Particles are dispersed into Milli-Q water externally by mixing. Suspensions are carefully sealed in a fluid cell (of 88 ml) made acoustically transparent, with circular holes on each sides covered with stretched Parafilm films. Mixtures are then subjected to successive focused acoustic pulses (3 cycles) of 20 different amplitudes with a randomized order at a fixed repetition period of 0.5 s.

Fig. \ref{FIG1}a) shows our experimental setup. Pressure waves are generated using a High-Intensity Focused Ultrasound, HIFU (Sonic Concepts H101, 1.1 MHz). Minimum pressures at the focal point are measured using a needle hydrophone (Onda HNR-0500) and are varied from 0.04 to -1.95 MPa. Cavitation detection is based on the backscattered sound of the nucleated bubbles, which act as strong acoustic scatterers \cite{Roy1990, Herbert2006}. Echo signals are detected by monitoring the HIFU transducer voltage between 0.08 and 0.1 ms after the excitation, corresponding to the time required by a wave to travel to the focal region and to come back to the transducer (see Fig. \ref{FIG1}c). Synchronized high-speed images show a clear correlation between the nucleated bubble positions and the backscattered signals, demonstrating the significant sensitivity of this passive technique, which has been used in a similar system \cite{Herbert2006}. For each pulse, a bandpass filter centred at 1.1 MHz is applied to the echo signal, and the resulting amplitude is compared to the noise level to detect cavitation events. Finally, cavitation probability is extracted based on 50 pulses of each amplitude and monitored over several hours. The suspension is mixed homogeneously using a homemade stirrer (at 800 RPM) and PTFE magnetic bar during the entire experiment.

\begin{figure*}[ht]
\begin{center}
  \includegraphics[width = 15 cm]{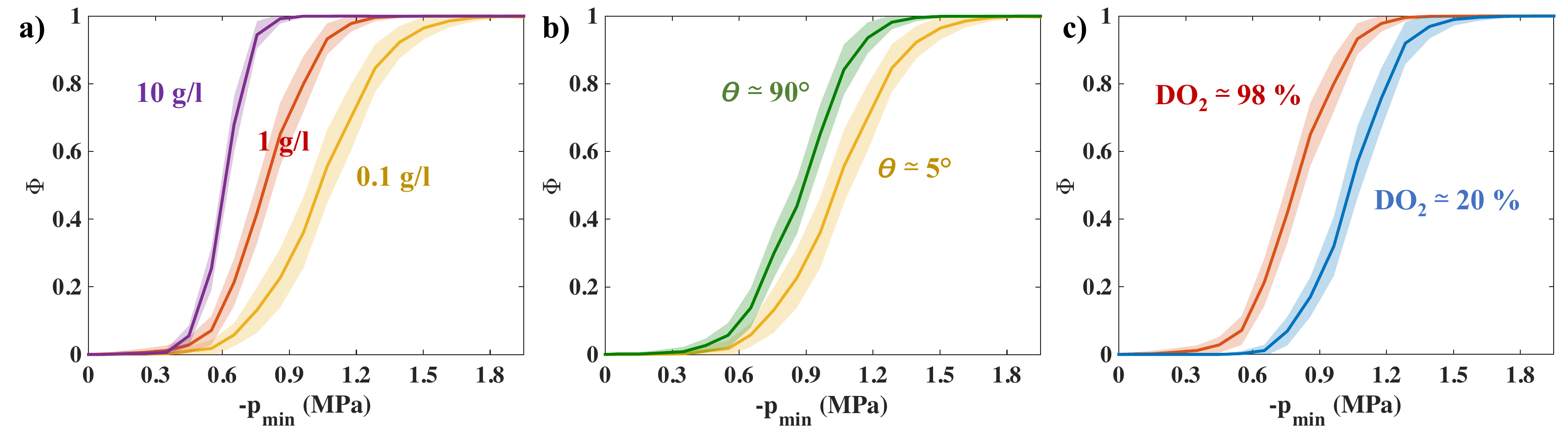}
\caption{\label{FIG3}{Stable cavitation probability curves ($\Phi$) for suspensions of different a) particle concentration, b) hydrophobicity where $\theta$ is the contact angle and c) amount of dissolved gas quantified by the level of dissolved oxygen (DO$_2$). Each solid line (shaded area) represents the average (standard deviation) of at least two separated suspensions measured in the same conditions over 4 hours with a 0.5 s pulse period ($\approx$ 28000 pulses per experiments).}}
\end{center}
\end{figure*}

Cavitation probability ($\Phi$) curves are plotted for thousand pulses in term of $p_{min}$, the minimum pressure reached at the focal point during one pulse (see Fig.~\ref{FIG1}d inset). Time evolution of the cavitation is analyzed using the critical pressure $p_{50}$, at which 50$\%$ of the pulse lead to cavitation, extracted by fitting the probability curves $\Phi$ with a cumulative normal distribution function $(1/2)\left(1+\text{erf}\left((p_{\text{min}}-p_{50})/(\sqrt{2}\sigma)\right)\right)$, as illustrated in Fig.~\ref{FIG1}d inset.
Typical evolution of this critical pressure $p_{50}$ over time for four independent experiments with a concentration of 1 g/l is shown in Fig.~\ref{FIG1}d. During the first 30 minutes of excitation (corresponding to 6000 pulses), the critical pressure evolves with time, and different trends are observed initially, suggesting different initial nuclei distributions. Interestingly, after 30 min the $p_{50}$ of all the mixtures stabilizes at a statistically reproducible value, which lasts for hours. Note that the stable behavior arises after thousands pulses (30 min), a regime not probed by previous studies reporting a nuclei deactivation \cite{Borkent2009} or a decrease of the cavitation activity \cite{Borkent2007}. In the following, we focus on this unexplored long-time behavior where the observed cavitation is independent of the initial conditions (i.e., statistically reproducible results for independent experiments).

The long-term cavitation stabilization observed suggests that the nuclei population of the water/particle mixture is able to reach a steady state after a certain amount of time. To get a deeper insight into this peculiar state, we explore the influence of the suspension parameters, such as the particle concentration, hydrophobicity, and the amount of dissolved gas content, known to impact the cavitation. The particle hydrophobicity was modified by coating a layer of dimethyldichlorosilane (DMDCS) by dip coating leading to a contact angle around 90$^{\circ}$ on smooth surfaces. Dissolved gases were depleted by placing the suspension in a vacuum chamber for one hour prior to the experiments, and the dissolved oxygen content was measured using an oxygen meter (TPS 90FL-T). Under these different conditions, similar time-dependencies are observed, and a stable later-time cavitation dynamic emerges after a similar timescale (30 min). However, the long-time behavior is found to depend upon all the parameters tested as shown in Fig. \ref{FIG3}. Note that each cavitation probability is based at least on two independent suspensions for experiment lasting for four hours with a 0.5 s repetition period.  The solid lines (resp. shaded area) represent the average (resp. standard deviation) of the cavitation probability over at least 50,000 pulses, or 2500 pulses per amplitude. 

An increase of the particle concentration results in a steepening of the long-time cavitation probability (shown in Fig.~\ref{FIG3}a), and the mixture of higher concentration reaches 100\% of cavitation at lower pressure excitation. However, the first cavitation events are detected at a similar tensile stress (-0.4 MPa) regardless of the particle concentration, suggesting that the easiest nucleus to cavitate are independent of the number of particles. Our results are in agreement with previous short-time study \cite{Roy1990} and imply the increase of the probability of having a nucleus in the HIFU focal region as the particle number increases.
Such influence of the particle concentration evidences that the cavitation observed at later-time originates from nuclei trapped in particle defects.

As shown in Fig.~\ref{FIG3}b, the particle hydrophobicity also impacts the long-time cavitation probability. Suspensions of 0.1 g/l hydrophobic particles exhibit a steeper and slightly shifted (toward lower tensile stress) cavitation probability compared to hydrophilic ones. This result indicates that the particle contact angle does not drastically change the easiest nucleus to cavitate (as shown by similar cavitation inception pressures observed) but increases the number of nuclei showed by the steeper probability. The nucleus stability increases with the contact angle in a later-time regime, similar to previous early-time experiments \cite{Marshall2003,Borkent2007,Belova2013}, supporting a nucleus origin for the observed cavitation.

Revealed in Fig.~\ref{FIG3}c is the influence of the dissolved gas on the cavitation. A decrease of the dissolved gas amount shifts the cavitation probability toward higher tensile stresses with no effect on the steepness, which suggests a shift of the nuclei distribution toward smaller sizes exhibiting higher stability to tensile stresses. This long-time tendency is similar to the one predicted by the crevice model and verified by previous experiments performed at short time, where a decrease of the gas content is found to increase cavitation threshold \cite{Strasberg1959,Crum1979} or to reduce cavitation activity \cite{Belova2013}. Moreover, this result evidences the nuclei, supporting the stable later-time cavitation, are of gaseous origin.

\begin{figure}
\begin{center}
    \includegraphics[width = 7 cm]{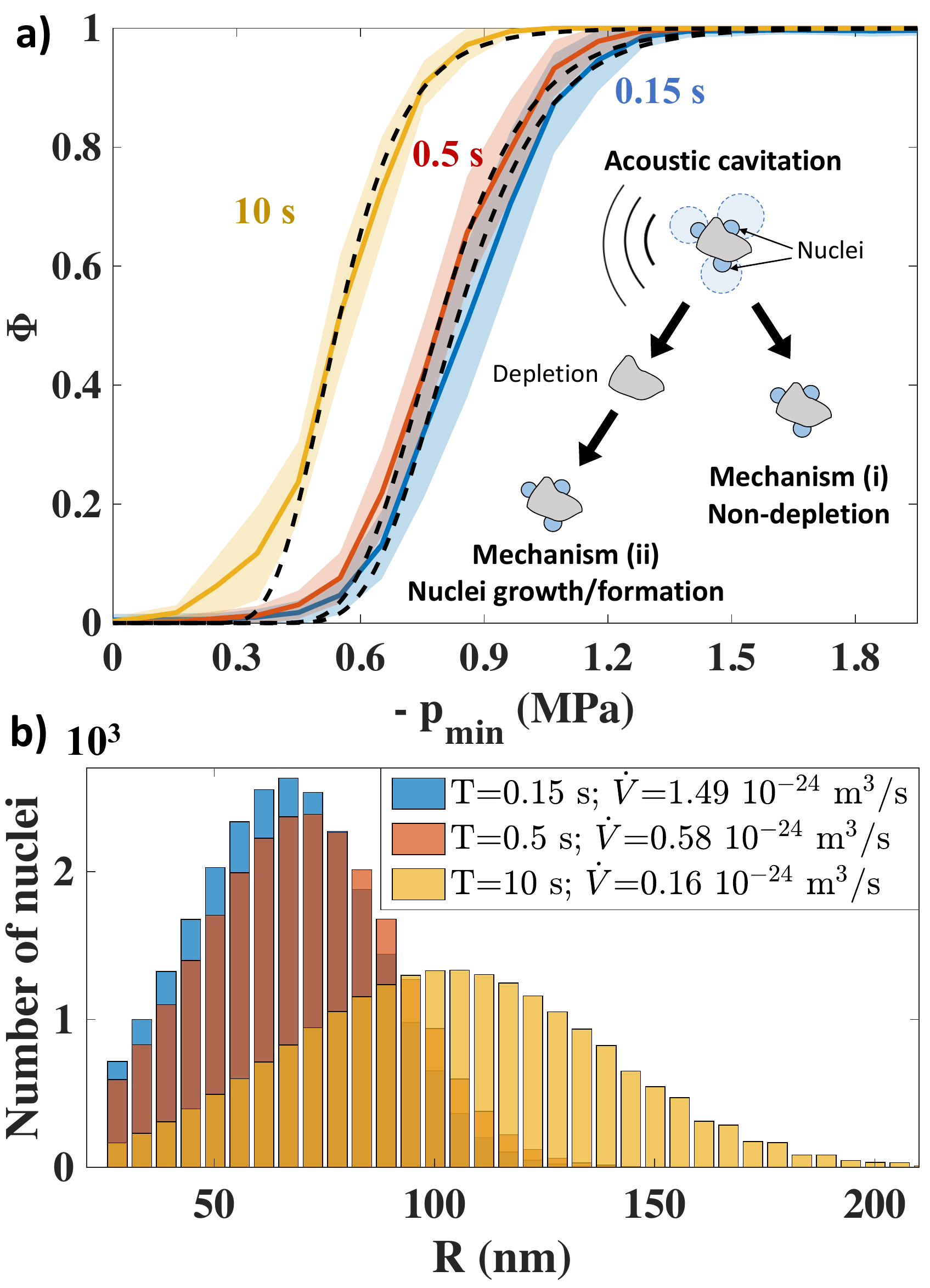}
\caption{\label{FIG4}{a) Stable cavitation probability ($\Phi$) of 1 g/l suspensions, measured after the initial state, for different pulses periods of $T$ = 0.15, 0.5 and 10 s. Solid lines and shaded areas represent the average and standard deviation, respectively, over at least 8000 pulses and two independent suspensions. Dashed lines are the cavitation probabilities obtained from the model developed with a constant nuclei growth rate (using Eq.~\eqref{eq:dist_grow} and \eqref{eq:prob_cav}). Inset: sketch of two possible mechanisms responsible for the long-term cavitation stabilization (i.e., stable nuclei population).} b) Nuclei distributions corresponding to the best fits shown in (a) for different $T$.}
\end{center}
\end{figure}

Cavitation nuclei have been found to be depleted under successive acoustic pulses \cite{Borkent2007,Borkent2009}, and if so the cavitation probability would have continuously drifted toward higher tensile stress in our experiments. Therefore, to sustain the observed stable cavitation (stable nuclei population), either (i) some nuclei exhibit a high stability and do not deplete over time or (ii) a continuous nuclei creation/growth is able to balance the depletion (see Fig.~\ref{FIG4}a inset). These two possible mechanisms can be distinguished by varying the excitation rate since an inexhaustible nuclei population (mechanism i) is expected to be independent of the depletion rate (as an intrinsic property of the suspension), and thus the resulting cavitation probability would be unaltered.

Under different excitation rates, i.e., using different repetition periods, $T$, the cavitation data ($p_{50}$) of a suspension still converges toward a stable and reproducible regime after few thousand pulses. The resulting long-time cavitation probabilities $\Phi$ of 1 g/l mixture are reported in Fig.~\ref{FIG4}. $\Phi$ is found to shift toward lower tensile stresses as $T$ is increased, without obvious change in the steepness. Therefore, the observed cavitation stabilization do not originate from highly stable nuclei (mechanism i) but result from a balance between nuclei deactivation and a growth or nucleation of nuclei (mechanism ii). Moreover, longer time between pulses (e.g., $T = $10 s in Fig.~\ref{FIG4}) leads to a higher cavitation activity, implying a spontaneous nuclei regeneration, i.e., letting the system at ``rest" increases the size and population of nuclei.

To decipher this self-regeneration mechanism, we consider different types of hydrophobic defects hosting the nuclei. Surface defects can be classified in two categories: narrow and wide crevices, based on their ability to stabilize a vapor bubble \cite{Atchley1989,SM}. On one hand, narrow crevices are highly confined and hydrophobic, vapor nucleation inside is spontaneous (without energy barrier) \cite{Giacomello2013}. Once formed, bubbles grow by gas diffusion \cite{SM}. Such nuclei regenerate in the order of 1 ms \cite{SM} $\ll T$, the pulse repetition period, thereby suggesting no influence of $T$, different from our experimental data. Narrow crevices are therefore in a negligible amount in the probed pressure range. On the other hand, vapor bubble cannot nucleate in wide crevices since no mechanical equilibrium exists \cite{Giacomello2013,SM}. Yet, in presence of dissolved gas, molecule of gas can adsorb on hydrophobic surfaces \cite{Fang2016,Schlesinger2018}, resulting in a local gas supersaturation required for nuclei equilibrium \cite{SM} and nucleation \cite{Wilt1986,Zhou2012} in wide crevices. Nucleated gas bubble can then grow by diffusion from the surface adsorbed gas layer to the bubble \cite{Yang2007}. Dissolved gas adsorption on surface, also assumed to be the origin of surface nanobubble formation \cite{Zhou2012,Seddon2011,Lohse2015,Yang2007,Fang2016}, occurs over a characteristic timescale of 10 minutes \cite{Yang2007,Fang2016}.

To shed light on the regeneration mechanism, we develop a model to further quantify the nuclei regeneration by estimating the nuclei distribution (hosted in the wide variety and number of particle defects) in our experiments, based on the stable cavitation probability $\Phi$. Consider a nuclei distribution made of $I$ classes, where each class is associated to a cavitation threshold $p_i$ having a population $N_i$. Under an acoustic pulse with a minimum pressure $p_{k}$, nuclei located in a volume $V_{ac}(p<p_i,p_{k})$, where the pressure $p$ induced by the acoustic excitation is lower than the threshold $p_i$, cavitate and collapse. The number of nuclei destroyed by the impulsion $p_k$ is then $N_iA_{i,k}$, where $A_{i,k}=V_{ac}(p<p_i,p_k)/V_s$ and $V_s$ the total volume of the mixture, and the average depletion over the $N_{ex}$ impulsions is given by $\alpha_i=\left(1/N_{\text{ex}}\right)\sum_{k=1}^{N_{ex}}A_{i,k}$. The volume $V_{ac}$ has been calibrated by mapping the acoustic field \cite{SM}. Note that we generalized, here, the depletion model previously proposed for one nucleus type excited by constant-intensity pulses \cite{Borkent2007} to a nuclei distribution subjected to different pressure pulses.

Nuclei regeneration is composed of a nucleation followed by a growth. As discussed in the SM \cite{SM}, we consider here that growth is the limiting process in nuclei regeneration as it provides self-consistent results. However, a more quantitative model would need to include the nucleation to fully capture nuclei evolution. Here, we model nuclei growth by introducing a factor $g_i$, which represents the proportion of nuclei growing from the class $i$ to the class $i+1$ between the impulsion $n$ and $n+1$. Combining the growth and the depletion, the evolution of the distribution is given by $N_i^{n+1} = N_i^{n}-\alpha_i N_i^{n} + g_{i-1} N_{i-1}^{n} - g_i N_i^n$. After few thousand pulses, the nuclei distribution reaches a steady state shown by a stable $\Phi$, i.e., $N_i^{n+1}=N_i^n=N_i$, leading to
\vspace{-0.1in}
\begin{equation}
    N_i=N_{i-1}\frac{g_{i-1}}{\alpha_i+g_i}.
    \label{eq:dist_grow}
    \vspace{-0.05in}
\end{equation}

The determination of $g_i$ and its evolution with $p_i$ is detailed in the SM \cite{SM}. In short, we assume nuclei growth occurs at a constant volume growth rate, $\dot{V}$, which can be seen as the average growth rate over the different sizes. Nuclei volumes, $V$, are expressed in term of cavitation threshold $p_i$, by assuming spherical nuclei, and the growth factor is then linked to the growth rate $\dot{V}$ and $p_i$ by considering nuclei residence times in each class. The growth factor decreases as bubble grow (increase of the residence time as $|p_i|$ decreases) and is given by $g_i=bp_i^4$, where $b$ is a constant containing $\dot{V}$. Finally, nuclei distribution is fully determined by assuming $N_0$ and $b$, through Eq.~\eqref{eq:dist_grow} \cite{SM}.

One can now link this distribution to our experiments by noticing that the probability of cavitation is the probability of having at least one nucleus in the focal volume with a threshold higher than the local pressure \cite{Messino1963}. If we assume, for clarity, the distribution to have only one class ($N,p_{th}$), the probability of not having one marked nucleus in the focal volume $V_{ac}(p<p_{th},p_k)$ during an impulsion $p_k$ is $1-V_{ac}/V_s$. The probability that none of the nuclei are in the focal region is  $(1-V_{ac}/V_s)^N$, and the probability that at least one of the nuclei is in the focal region, i.e., the cavitation probability is $\Phi=1-(1-V_{ac}/V_s)^N$ \cite{Messino1963}. The cavitation probability is generalised to a nuclei distribution by \cite{Messino1963,Herbert2006,Gateau2013}:
\vspace{-0.1in}
\begin{equation}
   \Phi(p_k)=1-\prod\limits_{i=1}^{I}\left(1-A_{i,k}\right)^{N_i}.
   \label{eq:prob_cav}
\vspace{-0.1in}
\end{equation}

Using Eq.~\eqref{eq:dist_grow} and \eqref{eq:prob_cav} and $N_0$ and $b$ as fitting parameters, experimental probabilities are fitted allowing the estimation of the growth rate given by $\dot{V}=3a\Delta p b/T$ \cite{SM}. The resultant best fits and their distributions are shown in Fig.~\ref{FIG4}. The predictions of $\Phi$ by this model agree with the experimental data for different $T$, albeit slight deviation at small excitation pressure. This deviation likely comes from the assumption on the spherical-shaped nuclei and on the constant volume growth rate. As $T$ increases, nuclei distribution widens and is shifted toward higher nucleus size. It should be note that the total number of nuclei is nearly constant $N_t\approx2.3\times10^4$ for all repetition periods. This indicates that surface defects are in a reproducible amount and that nuclei grow in the same defects and is consistent with a nuclei growth-limited regeneration. For $T=0.5$ s, nuclei corresponding to 50\% of cavitation have a volume $V_c = -a/p_{50}^3 \approx 1.4 \times 10^{-21}$ m$^3$ \cite{SM}, and the characteristic regeneration time is $\tau = V_c/\dot{V} \approx 40$ min. This result suggests that the nuclei regeneration is indeed driven by dissolved gas adsorption on surface. Note, this timescale is also consistent with the time required to reach a stable cavitation probability, revealed by Fig.~\ref{FIG1}d. Finally, the average growth rate $\dot{V}$ is found to decrease with $T$, shown in Fig.~\ref{FIG4}b, in agreement with the decay of the adsorption rate with time reported previously \cite{Yang2007}.

In summary, surprising and yet reproducible emergence of a stable cavitation probability is experimentally found for a particle/water mixture after a long-term excitation using several thousands of acoustic pulses. This stable cavitation was found to originate from a balance between nuclei deactivation by acoustic cavitation and a spontaneous regeneration of nuclei. These results highlight the dynamic character of the nuclei population, in opposition with the static view adopted up to now. The characteristic timescale of nuclei regeneration is in the order of tens minutes, suggesting a regeneration supported by dissolved gas adsorption on surfaces, similar to nanobubble formation. The method developed here provides a new tool to probe nanosize bubble formation and distribution on surfaces.

\begin{acknowledgments}
The authors gratefully acknowledge inspiring discussions with John Ralston (U South Australia), R\'egis Wunenburger (Sorbonne U), Detlef Lohse (U Twente) and Benjamin Dollet (U Grenoble-Alpes). This work was supported by the Collaborative Research and Development (CRD) of the Natural Sciences and Engineering Research Council of Canada (NSERC) and the Canadian Centre for Clean Coal/Carbon and Mineral Processing Technologies (C5MPT).
\end{acknowledgments}

Email addresses: {A.B. (bussonni@ualberta.ca)}; {P.A.T. (peichun@ualberta.ca)}; {Q.L. (qingxia2@ualberta.ca)}. 

\bibliographystyle{aipnum4-1}

\bibliography{biblio}

\pagebreak

\onecolumngrid
\begin{center}
  \textbf{\large Supplemental Material: \\``Cavitation nuclei regeneration in water-particle suspension"}\\[1cm]
\end{center}
\twocolumngrid

\setcounter{equation}{0}
\setcounter{figure}{0}
\setcounter{table}{0}
\setcounter{page}{1}
\renewcommand{\theequation}{S\arabic{equation}}
\renewcommand{\thefigure}{S\arabic{figure}}

\section{I. Narrow and wide crevices as nucleation sites}

\begin{figure}[htp!]
\begin{center}
  \includegraphics[width = 8.6 cm]{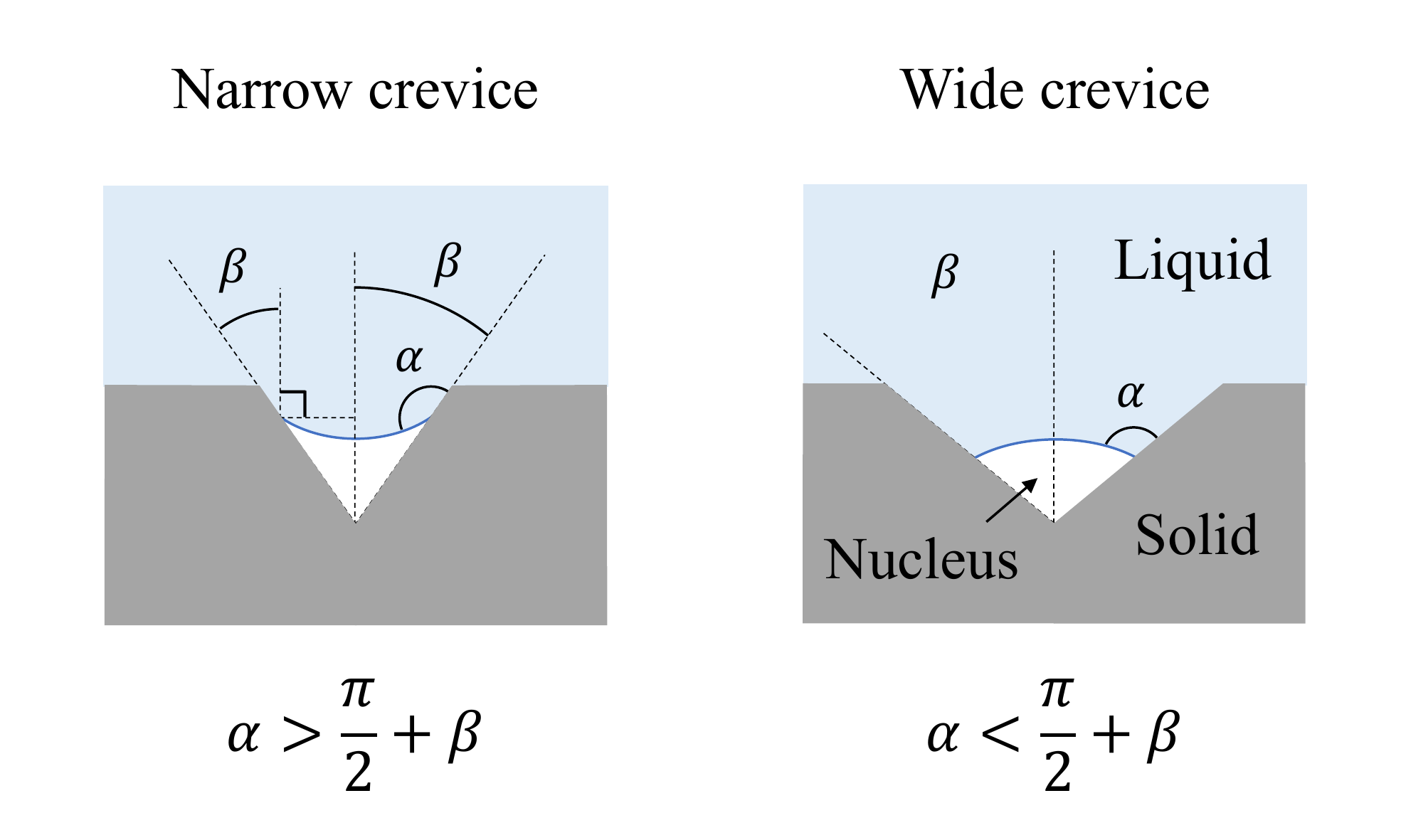}
\caption{\label{SI_crevices} Schematics of narrow and wide crevices.}
\end{center}
\end{figure}

Surface defects are commonly model as conical pits \cite{Atchley1989} with a cone half-angle $\beta$, and the hydrophobicity is characterized by the contact angle $\alpha$ (see Fig. \ref{SI_crevices}). Crevices are categorized based on their ability to stabilize a vapor bubble. Pressure in such bubble is equal to the vapor pressure $p_v = 2.3$ kPa, which is lower than the atmospheric pressure $p_0$. The mechanical equilibrium of the interface requires a force balance through pressure difference and the Laplace pressure across the interface:
\begin{gather}
    p_v = p_0 + \sigma\kappa,
\end{gather}
with the surface tension, $\sigma$, and the mean interface curvature, $\kappa$. Since $p_v < p_0$ the curvature $\kappa$ must be negative for a vapor bubble to be stable. Therefore, crevices that are able to stabilize a vapor bubble are defects sustaining negative curvature of the nucleus interface.

Using the simplified geometry shown in Fig.~\ref{SI_crevices}, this curvature requirement imposes a condition on the contact angle, $\alpha$, and the cone angle, $\beta$. If the contact angle is high and/or the cone angle is low such that $\alpha > \pi/2 + \beta$, crevices can host a stable vapor bubble and are called narrow crevices. On the one hand, for such narrow crevice, there is no energy barrier to nucleate a vapor bubble from a defect filled with water, i.e., the transition from the (completely-wetting) Wenzel to (gas-trapping) Cassie-Baxter state and vapor nuclei spontaneously nucleate in narrow crevices \cite{Giacomello2013}.  On the other hand, in wide crevices, if $\alpha < \pi/2 +\beta$, no mechanical equilibrium exists for a vapor bubble, thereby preventing any nucleation of vapor nuclei in wide defects, i.e., the Cassie-Baxter state with vapor bubbles does not correspond to a minimum of the grand potential for a liquid at atmospheric pressure \cite{Giacomello2013}.

In wide crevices, a mechanical equilibrium exists only if the pressure inside the nuclei is higher than the liquid pressure to balance the Laplace pressure, which is imposed by the positive curvature. In the presence of dissolved gas, nuclei are made of vapor and gas, and the bubble pressure $p_b = p_g + p_v$ can balance the Laplace pressure imposed by the positive curvature when the gas pressure $p_g > p_0-p_v$. In addition, the chemical potential equilibrium (i.e., Henry's law) imposes a dissolved gas concentration at the interface $c_b = H p_g > H (p_0 -p_v)$, with $H$ the Henry constant, which is higher than that in the bulk liquid given by $c_{\infty} = H(p_0-p_v)$ for a saturated solution (i.e., equilibrium with a flat interface). Nuclei equilibrium and nucleation in wide crevices, therefore, require gas supersaturation close to the interface, which could be induced in our experiments through gas adsorption onto hydrophobic defects \cite{Zhou2012,Yang2007}.

\section{II. Nucleus growth in narrow crevices}

Once a bubble nucleates in a narrow crevice, its interface, in addition to the mechanical equilibrium, must satisfy the chemical potential equilibrium, i.e., Henry's law. As nucleated bubbles contains only few gas molecules, the concentration $c_b \approx 0$, and dissolved gas starts to diffuse toward the nuclei. For narrow crevices, interface can slide toward the crevice mouth while keeping a negative curvature \footnote{In presence of a contact angle hysteresis, the condition for the interface to slide is $\alpha_r > \pi/2+\beta$, where $\alpha_r$ is the receding angle}. From the mechanical equilibrium, the bubble gas pressure is $p_b = p_0 + \sigma\kappa - p_v$, and the gas concentration at the interface $c_b$ remains smaller ($\kappa<0$) than the concentration far away in the liquid $c_{\infty} = H(p_0-p_v)$. The nucleus grows via dissolved gas diffusion and continues until the interface slides up to the crevice mouth, where it pins and becomes flat. Nucleus at this point is stable and fully reformed.

This growth by gas diffusion takes place over a characteristic timescale $t_d\sim\rho {r_0}^2/(D\Delta c)\sim 1$ ms \cite{Epstein1950}, where $\rho\lesssim1.2$ kg/m$^3$ is the air density, $r_0\approx 100$ nm is the characteristic defect size (corresponding to a cavitation threshold of $\sim$ -1 MPa observed in the experiments), $D\approx 2 \cdot 10^{-9}$ m$^2$/s is the diffusion coefficient of dissolved gas in water, and $\Delta c=c_{\infty}-c_b\approx c_{\infty}\approx 2 \cdot 10^{-2}$ kg/m$^3$ is the dissolved gas concentration gradient between the nuclei surface and the surrounding. Hence, if narrow crevices were the major defects supporting nuclei in the experiment, no dependencies of the cavitation probabilities on repetition period, $T$, would have been observed since it would have regenerated in few milliseconds. In short, with these reasons, the nuclei probed in our experiments are not supported by narrow crevices.

\section{III. Discussion on the nucleation}

As suggested by experiment results, nuclei depletion is balanced by a spontaneous regeneration mechanism, which is composed of a nucleation of nuclei followed by a growth. As indicated by the shift of the cavitation probability curves with different $T$ in Figure 3 of the main article, nuclei sizes increase with the repetition period, $T$. From this observation two cases can be considered: i) a nucleation-limited regime, where nuclei of bigger sizes are associated to a higher nucleation energy barrier, i.e., smaller nucleation rate, and are detectable only at long repetition period; ii) a growth-limited regime, where nuclei easily nucleate at small sizes and then grow toward bigger sizes over time. Note that in the latter case, the total number of nuclei is constant for different repetition periods, the distribution simply shifts toward bigger sizes. Here, we restrict our model to the growth-limited regime because it gives self-consistent results (of a constant number of nuclei for different repetition periods, $T$). We observe a deviation (of $\approx 15 \%$) between our experimental and theoretical results. Therefore, as a future study, a more quantitative model is required to consider not only growth but nucleation, which requires the detailed data of gaseous bubble nucleation barriers on solid surfaces, but such data are currently unavailable \cite{Bowers1996,Talanquer1995,Schmelzer2000,Lubetkin2003}.

Note that if the nucleation barriers and defect properties and number are known, nucleation can be incorporated in our model by adding a source term in the nuclei distribution equation: $N_i^{n+1}=N_i^{n}-\alpha_iN_i+g_{i-1}N_{i-1}-g_iN_i+J_iT$. $J_i$ is the nucleation rate (s$^{-1}$) of the nuclei of the class $i$. In a steady state, the nuclei distribution equation is:
\begin{equation}
N_i = N_{i-1}\frac{g_{i-1}}{g_i+\alpha_i}+\frac{J_iT}{g_i+\alpha_i}.
\end{equation}

\section{IV. Determination of the growth factor}

In our model, the nuclei growth is modeled by a growth factor $g_i = dN_{out}/N_i$, which characterize the proportion of nuclei growing from the class $i$ to $i+1$ ($dN_{out}$). Although dissolved gas adsorption has been experimentally evident \cite{Yang2007,Fang2016,Schlesinger2018}, gas adsorption kinetics and surface diffusion, which drive nuclei growth, are still lacking. To provide a simple estimation of the regeneration rate, we model the growth by a constant volume growth rate, which represents the average growth rate over different nuclei sizes (or classes, $i$).

To link the constant volume growth $\dot{V}$ to the nuclei distribution, we first express the cavitation threshold $p_i$ associated with each class to the nuclei volume. In the view of wide possible varieties and sizes of crevices in our experiments, as an approximation we consider the nuclei to be spherical, and the cavitation threshold of a nucleus with a radius $R_0$ is given by the Blake threshold \cite{Leighton2012}:
\begin{equation}
p_{th}=p_v-\left(\frac{32\sigma^3}{27\left(p_0-p_v+\frac{2\sigma}{R_0}\right)R_0^3}\right)^{\frac{1}{2}}.
\end{equation}

Typical cavitation threshold are measured in the order of 1 MPa. In this range, the initial pressure in the bubble is dominated by the capillary pressure $2\sigma/R_0\gg p_0-p_v$, and the cavitation threshold, $p_{th}$, is well approximated by:
\begin{equation}
    p_{th}=p_v-\left(\frac{32}{27}\right)^{\frac{1}{2}}\frac{\sigma}{R_0}\approx-\frac{0.77\sigma}{R_0}.
    \label{eq:cav_threshold}
\end{equation}

Note that the Blake threshold still holds for nucleus which destabilises inside a crevice, with the radius of curvature of the interface, $R_0$, which, in this case, depends on the crevice properties ($\alpha$ and $\beta$).

Nuclei of the class $i$ have a volume $V_i = (4/3) \pi R_i^3  = -a/p_i^3$, with $a = 4\pi/3 \cdot (0.77\sigma)^3\approx7.4\times10^{-4}$ m$^3 \cdot$Pa$^3$. A nucleus growing from the class $i$ to $i+1$ increases its volume by $\Delta V_i=3a\Delta p/p_i^4$, with $\Delta p=p_{i+1}-p_i$ constant for all the classes. For a constant volume growth rate $\dot{V}$, this increase takes place in a time $\Delta t_i=\Delta V_i/\dot{V}$, which correspond to the residence time of a nucleus in the class $i$. In a steady state, the number of bubble $dN_{in}$ entering the class $i$ at each impulsion is constant. After a time $\Delta t_i$, these nuclei grow to the class $i+1$, and $dN_{out}=dN_{in}$. The total number of nuclei in the class $i$ is the sum of bubble entering the class during the time $\Delta t_i$ and is given by $N_i = dN_{in}\Delta t_i/T$. Finally, the growth factor is expressed as :
\begin{equation}
g_i=\frac{dN_{out}}{N_i}=\frac{T}{\Delta t_i}=\frac{T\dot{V}}{3a\Delta p}p_i^4=bp_i^4,\label{S5}
\end{equation}
where $b$ is a constant and will be used as a fitting parameter. Once the data of $\Phi(p_k)$ fitted, the average volume growth rate is extracted using $\dot{V}=3a\Delta pb/T$.
\vspace{+0.2in}

\section{Acoustic calibration}

The minimum pressure generated by the transducer is first calibrated by placing a needle hydrophone (Onda HNR-0500) at the focal point. By varying the voltage over the entire range of the experiments, we experimentally determine the relation between the transducer voltage and the minimum pressure in the focal area $p_{focal}$.

\begin{figure}[t]
\begin{center}
  \includegraphics[width = 8.6 cm]{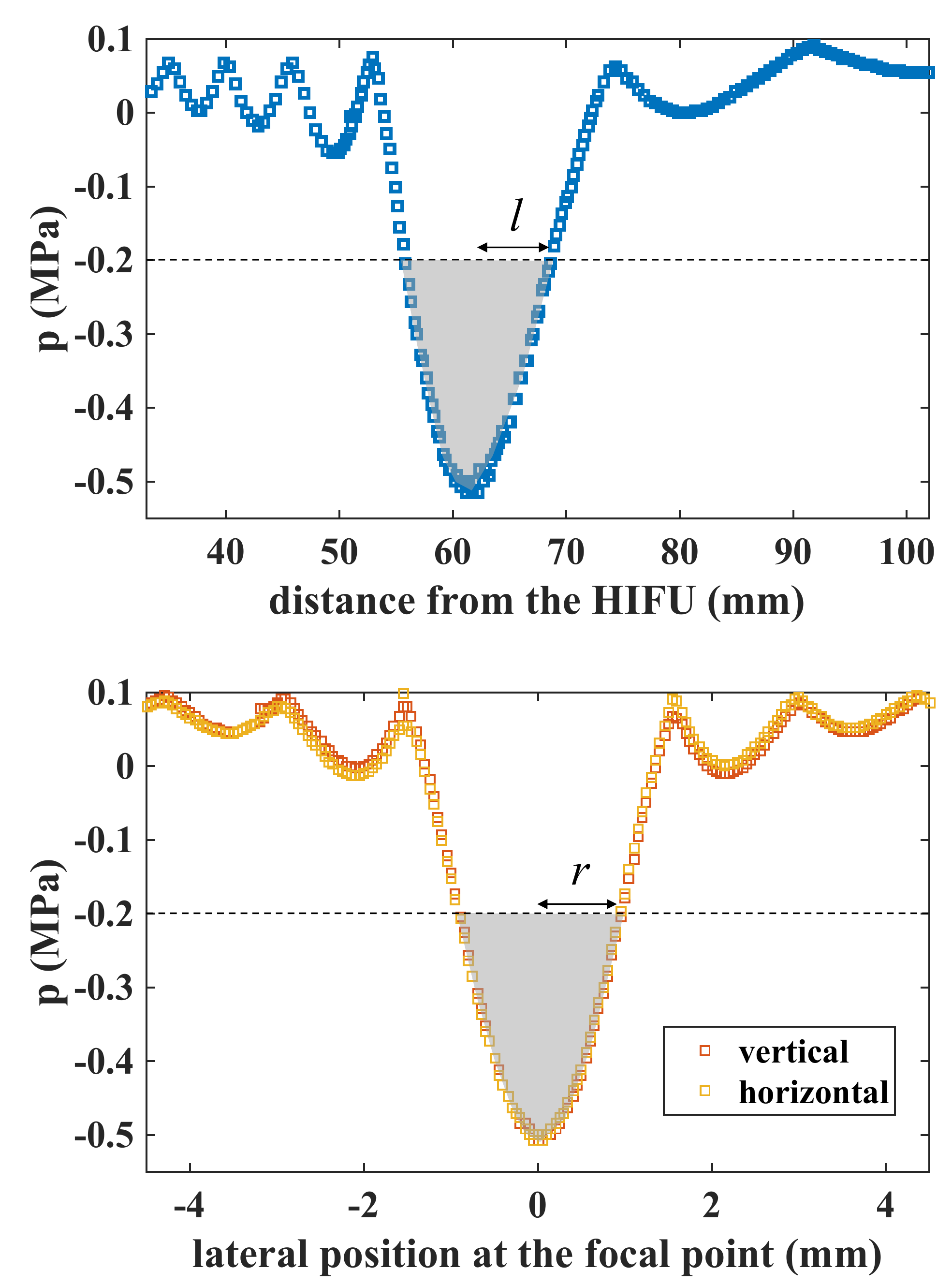}
\caption{\label{SI2}Minimum pressure of the acoustic field on the HIFU axe (top) and transverse axes at the focal position (bottom). The grey shaded area correspond to volume where the pressure is below -0.2 MPa and $l$ (resp. $r$) is the semi-major (resp. semi-equatorial) axis of the corresponding prolate spheroid.}
\end{center}
\end{figure}

The acoustic field is mapped with the same hydrophone at constant excitation ($p_{\text{{\it focal}}}\approx-0.5$ MPa) and low duty cycle; see Fig.~\ref{SI2}. The focal region have a prolate-spheroid shape with an axial axis $\approx 15$ mm and an equatorial diameter $\approx3$ mm. The volume where the pressure is below a value of $p_c$ (-0.2 MPa in Fig.~\ref{SI2}) for an excitation $p_\text{{\it focal}}=-0.5$ MPa, $V_{ac} (p < p_c, p_\text{{\it focal}})$, is represented in grey shaded area in Fig.~\ref{SI2}. To calibrate the acoustic volume for arbitrary $p_c$ and $p_{focal}$, we first neglect the nonlinear effects (e.g., transducer driven at low excitation) so the acoustic field, mapped in Figure \ref{SI2}, is proportional to the minimum pressure $p_{focal}$, previously calibrated. We then fit the pressure evolution in the focal area with the distance from the transducer (resp. the lateral distance) with a third (resp. fourth) degree polynomial to extract the axial length $l(p_c,p_{focal})$ and the equatorial radius $r(p_c,p_{focal})$ of the prolate-spheroid. Finally, the volume $V_{ac}(p_c,p_{focal})$ where the pressure field resulting from the excitation $p_{focal}$ is lower than $p_c$ is well approximated by:

\begin{gather}
    V_{ac}(p<p_c,p_{focal})\approx\frac{4\pi}{3}r^2l\\
    \approx0.079\left(\frac{p_c-p_{focal}}{p_{focal}-p_0}\right)^2-0.008\left(\frac{p_c-p_{focal}}{p_{focal}-p_0}\right)\label{S7}.
\end{gather}

The coefficient $A_{i,k}=V_{ac}(p<p_i,p_k)/V_s$ and $\alpha_i=\left(1/N_{\text{ex}}\right)\sum_{k=1}^{N_{ex}}A_{i,k}$ are calculated using Eq. (\ref{S7}) with $p_c=p_i$ and $p_{focal}=p_k$.

\section{Fitting procedure}

To fit the experimental data $\Phi(p_k)$ we used Eq. (2) of the main article on the cavitation probability:
\begin{gather}
    \Phi(p_k)=1-\prod\limits_{i=1}^{I}\left(1-A_{i,k}\right)^{N_i},
\end{gather}
where the nuclei distribution $N_i$ is calculated by combining Eq. (1) of the main article and Eq. (\ref{S5}) which leads to:
\begin{gather}
    N_i=N_{i-1}\frac{g_{i-1}}{\alpha_i+g_i}=N_0\prod\limits_{j=1}^{i}\frac{bp_{i-1}^4}{\alpha_i+bp_i^4}.
\end{gather}

Finally, the constant $N_0$ and $b$ are used as fitting parameters to adjust the experimental data $\Phi(p_k)$, and the volume growth rate $\dot{V}$ is extracted using the Eq. (\ref{S5}).
\newline 

Email addresses: {A.B. (bussonni@ualberta.ca)}; {Q.L. (qingxia2@ualberta.ca)}; {P.A.T. (peichun@ualberta.ca)}.

\end{document}